\documentclass[12pt,a4paper]{article}

\usepackage{epsfig}

\usepackage{amsmath,cite}

\def\e{\begin{equation}}
\def\f{\end{equation}}
\def\=#1{\overline{\overline #1}}

\def\_#1{{\bf #1}}

\def\.{\cdot}

\def\l#1{\label{eq:#1}}
\def\r#1{(\ref{eq:#1})}
\def\am{\left(\begin{array}{c}}
\def\amm{\left(\begin{array}{cc}}
\def\a{\end{array}\right)}

\title{On geometrical scaling of split-ring and double-bar resonators at optical frequencies}

\author{Sergei Tretyakov}

\date{Radio Laboratory / SMARAD Center of Excellence\\ TKK Helsinki University of
Technology \\P.O. Box 3000, FI-02015 TKK, Finland\\
}

\begin{document}
\maketitle {\center \large

Address for correspondence:

Sergei Tretyakov\\
Radio Laboratory, Helsinki University of Technology,\\
P.O. Box 3000, FI-02015 TKK, Finland.

Fax: +358-9-451-2152

E-mail: sergei.tretyakov@tkk.fi

}

\parskip 7pt

\vspace{0.5cm}
\begin{center}
\section*{Abstract}
\end{center}

In this paper we consider the resonant frequency of split-ring resonators and
double-bar resonators
used to create artificial magnetic response at terahertz and optical frequencies.
It is known that geometrical scaling of the resonant frequency of split rings
breaks down at high frequencies (in the visible) due to electromagnetic properties of
metals at those frequencies. Here we will discuss this phenomenon in terms of
equivalent inductance and capacitance of the ring, derive an approximate
formula for the resonant frequency of split rings in the visible and show how
the limiting value of the resonant frequency of extremely small split rings and double bars depends on
particle shape.

\textbf{Key words:} Negative index materials, negative permeability,
split ring, double-bar resonators, artificial magnetics.

\newpage

\section{Introduction}

Realization of low-loss metamaterials possessing properties of an artificial magnetic in the visible
would allow realization of artificial materials with
negative index of refraction and open a way for many novel applications.
One of the possible routes is the use of nano-sized resonant
metal particles of complex shapes. Various shapes have been proposed, and
the operational principle is basically the same: The particle shape is chosen so that one of the lowest eigenmodes
has a current distribution with a large magnetic moment. In particular, several variants of split rings
\cite{Linden,Moser,Kat} and double bars \cite{Kildishev,Shalaev} have been manufactured and measured in
frequency range of hundrends of teraherts and in the visible. Also,
arrays of pillar pairs have been studied in \cite{Grigorenko}.  However, it has been noticed
that the geometrical scaling of the particle resonant frequency breaks down when the
working frequency gets higher, so that the resonant frequency saturates at the level of
several hundreds of terahertz \cite{Zhou,Klein}. This effect is explained by the plasmonic behavior of
metals at optical frequencies, usually in terms of kinetic energy of electrons carrying current
and an additional ``electron self-inductance'' \cite{Zhou}.

In this paper we theoretically study the phenomenon of geometrical scaling breakdown
and saturation of the resonant frequency  using the equivalent circuit model.
We show that the full model of the phenomenon requires not only an additional inductance, but also an additional capacitance.
The model correctly gives the ultimate maximum resonant frequency which is equal to the plasma frequency of metal.
This model allows us to study the dependence of the achievable resonant frequency on the
particle shape.

\section{Magnetic mode resonant frequency in the visible}

\begin{figure}[h!]
\centering
\epsfig{file=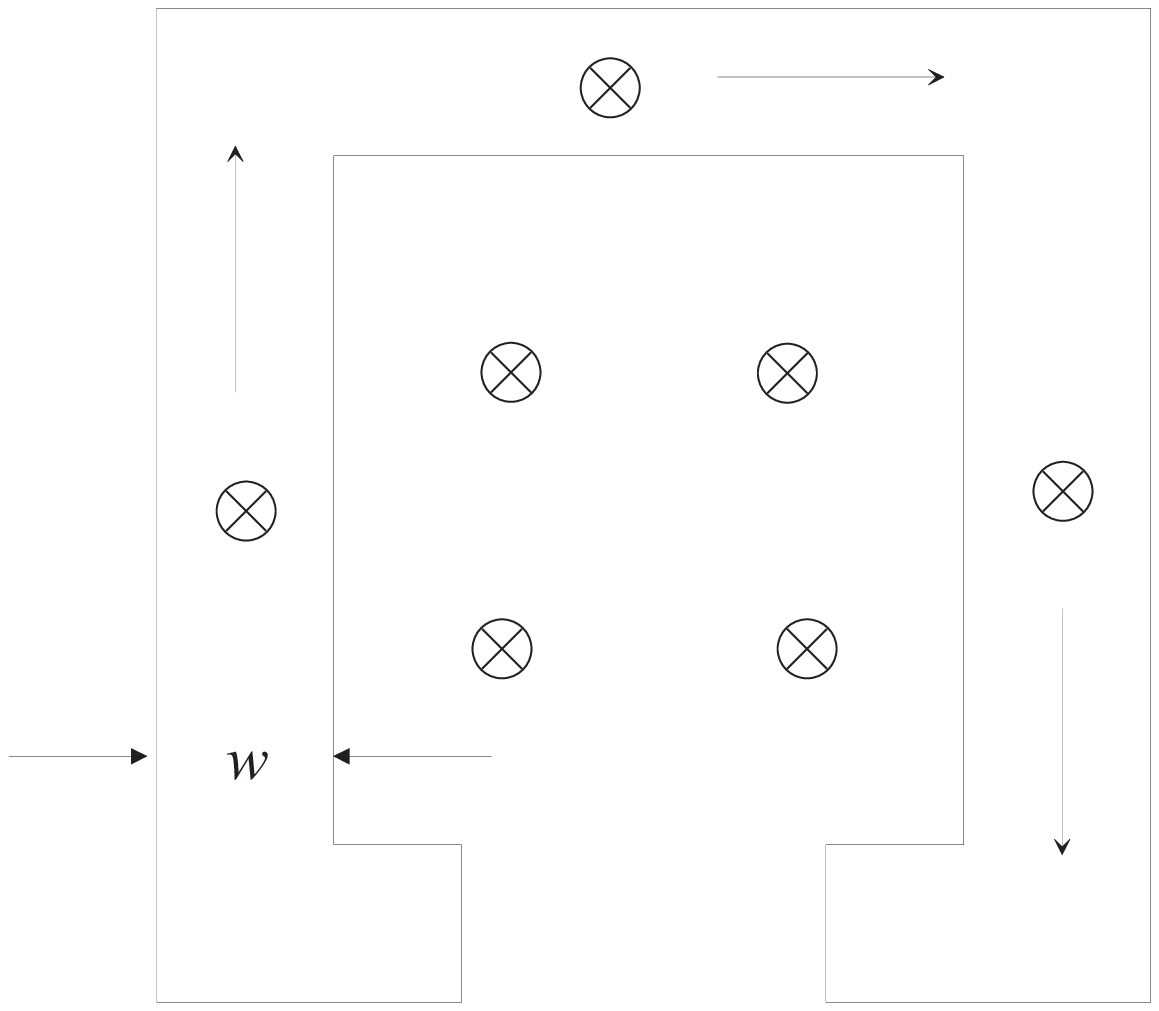,width=0.4\textwidth}\ \ \ \
\epsfig{file=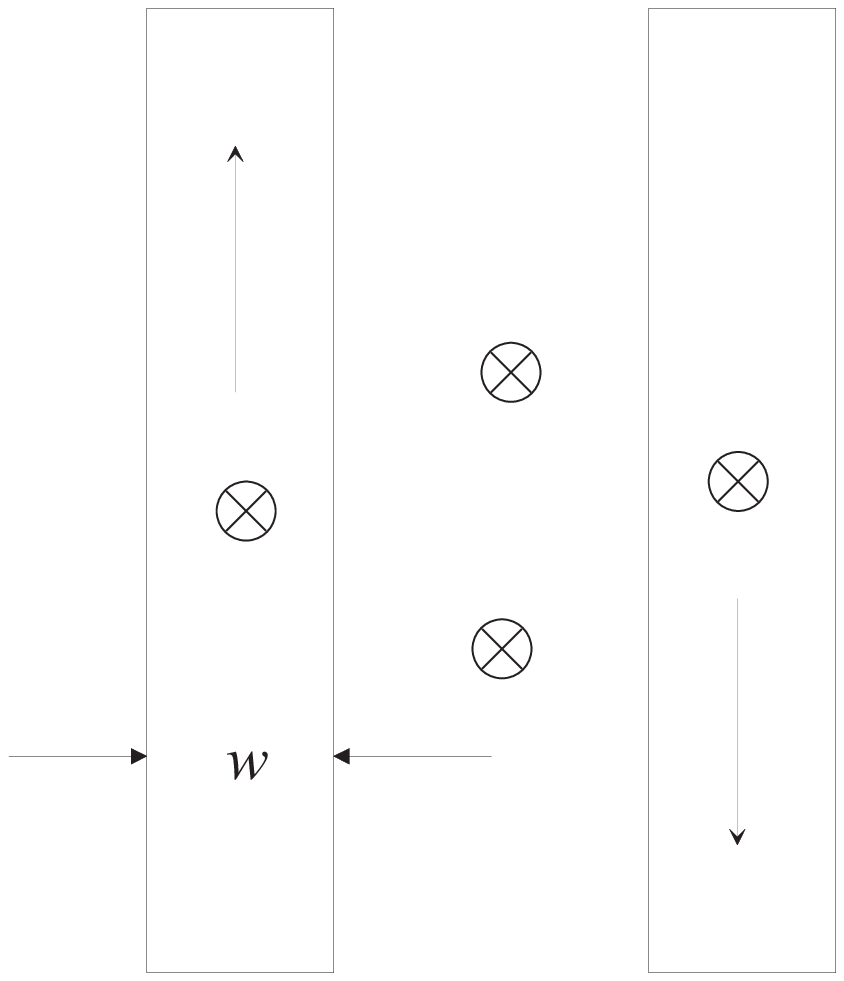,width=0.4\textwidth}
\caption{The geometry of split rings and double-bar particles. The particles are fabricated
from a strip of thickness $h$. The total length of metal strips in each particle is denoted by
$l$. Arrows show the current direction of the
magnetic mode. Magnetic flux and electric field exist not only in air but also inside metal strips.} \label{geometry}
\end{figure}

In contrast to the microwave regime, in the visible range metals do not behave as
ideal conductors, instead, their permittivity can be modeled by the Drude model:
\e \epsilon=\epsilon_0\left(1-{\omega_p^2\over \omega^2}\right)\f
where $\omega_p$ is the plasma frequency.
This means that the fields in the bulk of metal strips cannot be neglected or modeled as fields in a
thin skin layer. Let us assume that the current density is approximately uniform
over the cross section of the strip and express the displacement current density in the metal strip as
\e \_J_d= j\omega \_D=j\omega \epsilon_0\left(1-{\omega_p^2\over \omega^2}\right)\_E
\l{Drude}\f
Next, we can relate the total current passing the ring cross section with the voltage
along the ring perimeter, because the current amplitude equals $I=J_dwh$ (again assuming uniform current
distribution over the ring cross section) and the voltage is $V=El_{\rm eff}$. Here the effective
ring length can be approximated as $l_{\rm eff}=(\pi/2) l$, where $l$ is the physical length,
because the current distribution along an open loop or along short bars is approximately sinusoidal. The result reads:
\e I=J_dwh=j\omega \epsilon_0{wh\over l_{\rm eff}}\left(1-{\omega_p^2\over \omega^2}\right) V \f

This relation can be interpreted in terms of additional inductance and capacitance of the
ring due to magnetic and electric fields {\em inside} the metal ring:
\e I=\left(j\omega {\epsilon_0wh\over l_{\rm eff}}+{\epsilon_0wh\omega_p^2\over {j\omega l_{\rm eff}}}
\right)V=\left(j\omega C_{\rm add}+{1\over j\omega L_{\rm add}}
\right)V\f
The additional capacitance and inductance can be then identified from the above equation as
\e C_{\rm add}= {\epsilon_0wh\over l_{\rm eff}},\qquad L_{\rm add}={l_{\rm eff}\over
\epsilon_0 wh\omega_p^2}\l{LC_add}\f
Formula \r{LC_add} for $L_{\rm add}$ was derived earlier in \cite{Zhou} in terms of kinetic energy
of electrons in metal.

\begin{figure}[h!]
\centering \epsfig{file=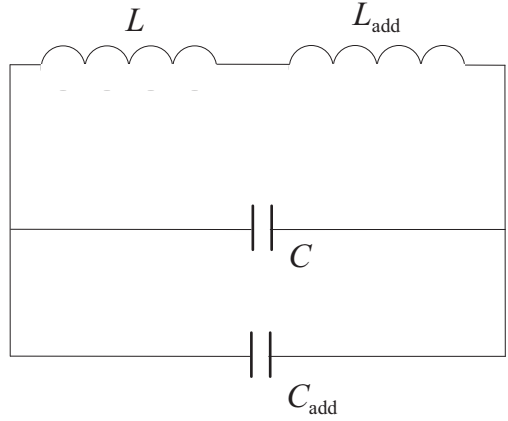,width=0.4\textwidth}
\caption{Equivalent circuit of a split ring or a double-bar particle in the visible.
$L_{\rm add}$ and $C_{\rm add}$ model additional stored energy inside the metal volume.} \label{circuit}
\end{figure}

The resonant frequency of the split ring is given by
\e \omega_0={1\over \sqrt{(L+L_{\rm add})(C+C_{\rm add})}}\l{simple}\f
which differs from the earlier model in \cite{Zhou} by the additional capacitance term.

The parameters of the ideally conducting particle $L$ and $C$ depend on
the dimensions and the shape of the particle. Considering the split-ring geometry, we can write:
\e L\approx \mu_0 {l\over
4}\ln {8l\over w+h},\qquad C\approx \epsilon_0 {wh\over \delta}\f where
we have approximated the split capacitance as the parallel-plate
capacitance ($\delta$ is the split width). For the inductance
estimations, we can use the formula for the round-loop inductance
\e L=\mu_0 a \ln{8a\over r_0}\f
and estimate the equivalent loop radius $a\approx l_{\rm eff}/(2\pi)=l/4$ and the
equivalent wire radius $r_0\approx (h+w)/4$. Finally, the resonant frequency of split rings reads
\e
\omega_0={1\over \sqrt{\left(1+{\pi\over 2}{l\over
\delta}\right)\left(\mu_0\epsilon_0 {wh\over 2\pi}\ln {8l\over w+h} +{1\over
\omega_p^2}\right)}} \l{rf}\f

Metal losses can be modeled by replacing formula \r{Drude} by
\e \_J_d= j\omega \epsilon_0\left(1-{\omega_p^2\over {\omega^2-j\omega\gamma}}\right)\_E\f
where $\gamma$ is the damping coefficient.
The result for the resonant frequency \r{simple} changes to
\e \omega_0=\sqrt{{1\over (L+L_{\rm add})(C+C_{\rm add})}-{R\over (L+L_{\rm add})^2}}\l{reson}\f
where the equivalent resistance reads
\e R={\gamma l_{\rm eff}\over \epsilon_0wh\omega_p^2}\f

\begin{figure}[h!]
\centering \epsfig{file=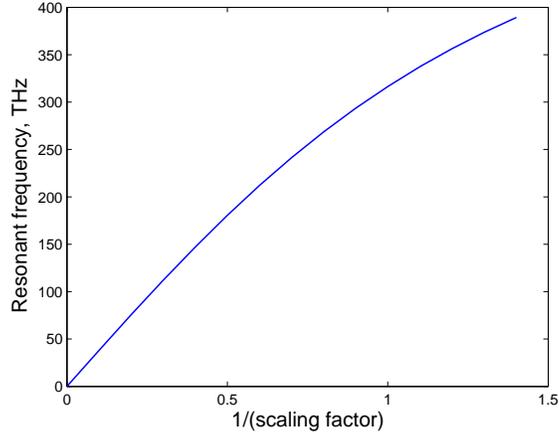,width=0.5\textwidth}
\caption{Resonant frequency of a split ring particle [formula \r{rf}] as a function of geometrical scaling factor.
Particle parameters are the same as in paper \cite{Klein}.} \label{scaling}
\end{figure}

A numerical example of the dependence of the resonant frequency on
the geometrical scaling factor is given in Figure~\ref{scaling}. The
particle is a split ring made of gold, and its dimensions are the
same as in paper \cite{Klein}. The plasma frequency of gold is
assumed to equal $\omega_p= 13.8\times 10^{15}$ 1/s, and the loss
factor $\gamma=107\times 10^{12}$ 1/s \cite{Ishi}. Despite several
model simplifications, the estimated resonant frequency (316.4 THz)
corresponds surprisingly well with numerical and experimental
results of \cite{Klein} (close to 330 THz). It is interesting to
note that metal losses have negligible effect on the resonant
frequency in this case. This can be understood by comparing the
values of $\omega_0 L_{\rm add}$ and $R$, which differ only by
replacement of the resonant frequency $\omega_0$ by the loss factor
$\gamma$. Because $\omega_0\gg\gamma$, the contribution of the term
proportional to $R$ in \r{reson} can be neglected. However, losses
degrade the resonator quality factor and one has to take them into
account especially at very high frequencies. In the optical region,
the level of losses depends not only on the material but also on the
particle shape, because of Landau damping (scattering on particle boundaries).
This means that the choice of the optimal shape should include
consideration of losses in addition to the resonant frequency of the particle.

\section{Discussion and conclusions}

In order to increase the resonant frequency, the denominator in
formula \r{rf} should be made as small as possible. Clearly, this
has a fundamental limit since even in the limit of zero dimensions the
limiting value is not infinity but equals to the plasma frequency of
the metal (it is stated also in paper \cite{Klein} that the limit is
determined by the plasma frequency). This is a very expected
conclusion because even if in the limit of small size the loop
closes completely and there is zero magnetic flux outside the
particle, electrons inside metal still can oscillate, and the
resonant frequency of these oscillations is the plasma frequency.
Inspecting formula \r{rf} we can see that the preferred design should
minimize the strip cross section area $wh$. This, however, has its
own limit determined by increased losses for smaller values of $wh$.
In addition, our results reveal that the limiting resonant frequency
for particles with all the dimensions going to zero proportionally
to the same scaling factor depends also on the particle shape:
\e
\lim\left\{\omega_0\right\} ={\omega_p\over \sqrt{1+{\pi\over
2}{l\over \delta}}}\f [see formula \r{rf}].
The ultimate highest resonant frequency equals to the plasma frequency of metal:
\e \max\{\omega_0\}={1\over \sqrt{C_{\rm add}L_{\rm add}}}=\omega_p\f
and it can be in principle achieved if in
diminishing the particle dimensions \e \lim \left\{{l\over
\delta}\right\}=0 \l{best}\f
This design recipe suggests that, when reducing the
particle size, the loop or bar length should be reduced in larger
proportion than the gap width. This has a clear meaning of reducing
both geometrical inductance and capacitance, since the inductance is
proportional to $l$, while the capacitance is inversely proportional to
$\delta$.

Finally, in choosing the particle shape it is important to remember that the current induced
by external magnetic field (in the LC-circuit model) is proportional to
$\omega^2 (C+C_{\rm add})/(1-\omega^2\omega_0^2)$. This means that comparing two particles with
the same resonant frequency $\omega_0$, the induced current is stronger in the particle with the larger
total capacitance. This consideration suggests to decrease capacitance in a smaller proportion than
inductance, if the goal is to design an optical metamaterial, but this contradicts with the criterium of
raising the limiting resonant frequency \r{best}. Thus, the particle shape choice is a compromise between the
increased resonant frequency and strong magnetic response.

\section*{Acknowledgements}
This work has been done within the frame of the European Network of Excellence {\itshape Metamorphose}. The
author wishes to thank Liisi Jylh\"a for helpful comments.

\end{document}